\documentclass[twocolumn,showpacs, aps, superscriptaddress, eqsecnum, prd, 
notitlepage, showkeys, nofootinbib, floatfix]{revtex4-2}
\usepackage{graphicx}
\usepackage{bm}
\usepackage{amsmath,amssymb}
\usepackage{natbib}
\usepackage{booktabs}
\usepackage{color}
\usepackage{units}
\usepackage{aas_macros}

\usepackage[normalem]{ulem}
\usepackage{orcidlink}

\hypersetup{colorlinks=true, citecolor=teal, urlcolor=teal, linkcolor=teal}

\usepackage{lineno}
\newcommand{\nsbh}{GW200105}
\newcommand{\gwecc}{\texttt{gw-eccentricity~}}
\newcommand{\seob}{\texttt{SEOBNRv5EHM~}}
\begin{document}

\title{
A universal framework to identify eccentric binary mergers: GW200105 case study}

\author{Teagan A.~Clarke~\orcidlink{0000-0002-6714-5429}} \email{teagan.clarke@princeton.edu}
\affiliation{Department of Physics, Princeton University, Princeton, New Jersey, 08544, USA}
\affiliation{School of Physics and Astronomy, Monash University, VIC 3800, Australia}
\affiliation{OzGrav: The ARC Centre of Excellence for Gravitational Wave Discovery, Clayton, VIC 3800, Australia}

\author{Isobel M.~Romero-Shaw~\orcidlink{0000-0002-4181-8090}} 
\affiliation{Gravity Exploration Institute, School of Physics and Astronomy, Cardiff University, Cardiff, CF24 3AA, United Kingdom}

\author{Charlie Hoy~\orcidlink{0000-0002-8843-6719}}
\affiliation{University of Portsmouth, Portsmouth, PO1 3FX, United Kingdom}

\author{Jakob Stegmann~\orcidlink{0000-0003-2340-8140}}
\affiliation{Max Planck Institute for Astrophysics, Karl-Schwarzschild-Str. 1, 85741 Garching, Germany}

\author{Paul D.~Lasky~\orcidlink{0000-0003-3763-1386}}
\affiliation{School of Physics and Astronomy, Monash University, VIC 3800, Australia}
\affiliation{OzGrav: The ARC Centre of Excellence for Gravitational Wave Discovery, Clayton, VIC 3800, Australia}

\author{Eric Thrane~\orcidlink{0000-0002-4418-3895}}
\affiliation{School of Physics and Astronomy, Monash University, VIC 3800, Australia}
\affiliation{OzGrav: The ARC Centre of Excellence for Gravitational Wave Discovery, Clayton, VIC 3800, Australia}

\date{\today}

\begin{abstract}

Orbital eccentricity in gravitational-wave signals from merging compact object binaries is a powerful indicator of their formation channel. 
Several binary black hole mergers and a neutron star--black hole merger have been reported to exhibit signs of eccentricity, but which events are identified and the significance of the eccentricity differs between studies. 
Measurements of eccentricity can change depending on the choice of prior.
The choice of prior is subtle: eccentricity is commonly measured at an arbitrary reference frequency, which varies from study to study.
We use the candidate eccentric neutron star--black hole merger GW200105\_162426 as a case study, employing a range of priors and reference frequencies, and find the results to be strongly prior-driven. 
We show that the varied results reported across different studies can be partially reconciled by accounting for the evolution of eccentricity with reference frequency. 
In order to make conclusive statements about eccentricity, we propose a detection statistic that does not depend on reference frequency, and which marginalises over astrophysically-motivated distributions in eccentricity. 
Using this detection statistic, we find reduced support for the eccentric hypothesis for GW200105\_162426: we obtain a natural log Bayes factor $\ln {\cal B} \leq {0.9}$ comparing the eccentric, aligned-spin hypothesis to the quasi-circular, precessing hypothesis.
Our results cast doubt on the eccentric interpretation of GW200105\_162426 and underscore the importance of modelling the astrophysical distributions of eccentricity in nature.

\end{abstract}

\maketitle

\section{Introduction}
\label{sec:intro}
Understanding the formation channels of merging compact object binaries observed by LIGO--Virgo--KAGRA (LVK) \citep{adv_ligo_2015, AdvancedVirgo, 2020_Kagra} remains a central problem in gravitational-wave astronomy.  
Measuring the orbital eccentricity, along with the component masses and spins, of the black holes in a binary can help determine how the binary formed. 
Inspiralling binaries efficiently circularise through the emission of gravitational waves \citep{Peters64}, so isolated binaries are expected to have quasi-circular orbits when they enter the LVK observing band at $\unit[\approx10]{Hz}$. Meanwhile, binaries formed through dynamical assembly can be driven to arbitrarily high eccentricities, and maintain measurable eccentricity when they enter the LVK band \citep[e.g.,][]{Oleary2009, Rodriguez18b, Zevin:2021:seleccentricity}. 

Signatures of dynamical assembly, such as orbital eccentricity \citep[e.g][]{Lower18, Romero-Shaw:2019itr, Romero-Shaw2021, OSheaKumar2021, romero_shaw_22, Iglesias2024, Gupte:2024:eccentricity} and misaligned spins \citep{GWTC-2_RnP, gwtc3, Hannam2022, Antonini:2025:hierarchical, Stegmann2025b, Abac2025}, inferred in some of the existing gravitational-wave observations, suggest that dynamically-formed systems may make up a substantial sub-population of binary black hole mergers.
Binaries in hierarchical field triples may also be driven to merge rapidly by the influence of the tertiary, which can drive up the binary's eccentricity through von Zeipel-Lidov-Kozai oscillations \citep{vonZeipel1910, Lidov62, Kozai62}, and can lead to measurable eccentricity at $\unit[10]{Hz}$ \citep[e.g.,][]{Naoz2013, Naoz_2016, RodriguezAntonini:2018:triples}.

One challenge associated with measuring eccentricity in gravitational-wave data is the relative lack of accurate, efficient eccentric waveform models. 
While none of the currently available waveform models include all of the required physics to accurately describe an eccentric inspiral, merger, and ringdown, this has been an area of rapid development and expansion. Several waveform models now include two variable eccentric parameters\footnote{An angular parameter, such as the mean anomaly, is required in addition to the orbital eccentricity to fully describe an eccentric orbit.} and higher modes \citep[e.g.,][]{Ramos-Buades2023, Gamboa2024, deLlucPlanas2025, Paul2025, clarke_2022}, and one inspiral-only model \citep{Morras2025_waveform} contains both eccentricity and spin-induced precession \citep{Apostolatos1994}. 

Despite the challenges in accurately modelling eccentricity in gravitational-wave inspirals, several binary black hole mergers in the current gravitational-wave transient catalogue show possible signs of eccentricity. These include GW190521 \citep{Gamba_2021, Romero-Shaw:2020:GW190521, Gayathri_2022} (although see also, e.g., Refs.~\citep{Ramos-Buades:2023:analysis, Gupte:2024:eccentricity}), GW190620\_030421 \citep{Romero-Shaw:2020:GW190521}, GW191109\_010717, GW200208\_222617~\citep{romero_shaw_22}, GW151226 and GW170608 \citep{Wu_2020, OSheaKumar2021}, GW190929 \citep{Iglesias2024}, GW200129 \citep{Gupte:2024:eccentricity, Tang2026} (although this event has known data quality issues that may affect eccentricity measurements \citep[e.g.,][]{Payne2022, Gupte:2024:eccentricity}), GW231001\_140220 \citep{Gupte2026, Malagon2026} and GW231123 \citep{Jan2025, Xu2026}. 
However, it is currently unclear whether these events are truly eccentric, or whether they would be better-explained by waveform models containing additional physics like spin-induced precession \citep[e.g.,][]{Romero-Shaw:2023:EccOrPrecc, Iglesias2024, Tibrewal2026}. In addition to hints of eccentricity among binary black hole mergers, Refs.~\cite{Fei2024} and \cite{Morras2025} report a non-zero measurement of eccentricity in the neutron star--black hole (NSBH)  merger GW200105\_162426 \citep{nsbh_detection} (hereafter referred to as \nsbh). Ref.~\cite{Morras2025} reports that the eccentricity of \nsbh~measured at a reference frequency of $\unit[20]{Hz}$ is $e_{20} \approx 0.15$ using the inspiral-only \texttt{pyEFPE} waveform model \citep{Morras2025_waveform}, which includes the effects of both eccentricity and spin-induced precession. 
These results have been supported in follow-up studies employing a variety of waveform models and inference techniques \citep[e.g.,][]{Planas2025, Tiwari2025, Kacanja2025, Phukon2025, Kacanja2025, Jan2026, Roy2026}.

Robustly measured eccentricity in an NSBH merger would provide particular insight into its astrophysical formation channel. While close few-body interactions in dense environments have been proposed as an origin for eccentric binary black hole mergers, the low mass ratio of NSBH mergers and large natal kicks at the neutron star formation typically inhibit efficient formation in these environments \citep[e.g.,][]{Clausen2013, Bae2014, Petrovich2017, Ye2020, Fragione2020}. 
Instead, Ref.~\citep{Stegmann2025} shows that the dynamics in hierarchical triples may naturally lead to a large fraction of NSBH mergers retaining non-negligible eccentricity in the LVK band.

However, the statistical significance of these claims is still ambiguous. Ref.~\cite{Morras2025} rules out $e_{20}<0.028$ with 99.5\% credibility. Given that the LVK collaboration has detected $\gtrsim 200$ events \citep{GWTC4}, it would not be surprising for at least one signal to be embedded in random noise with a fluctuation that mimics eccentricity with a chance probability of 0.5\%.
This raises the question: is \nsbh\ definitely eccentric, or is it merely the most eccentric-looking event in a large catalogue of circular binaries?

In this paper we use \nsbh~to examine the impact of prior choice and reference frequency on the results of eccentric gravitational-wave inference. 
In Section~\ref{sec:conversions}, we explore the evolution of eccentricity with frequency and the impact of this evolution on a uniform prior on eccentricity. 
In Section~\ref{sec:inference}, we show that the inferred eccentricity is extremely sensitive to the choice of prior and reference frequency, and that reasonable alternative choices yield far less confident eccentricity measurements. 
In order to address the issue of prior dependence, we propose an astrophysically-motivated \textit{eccentricity detection statistic} in Section~\ref{sec:detection_stat}.
The statistic employs theoretical models for the distribution of eccentricity for different astrophysical formation channels.
The resulting inference is still model-dependent, as the statistic depends on the astrophysical models of eccentricity; but the models are physically motivated, and the results do not depend on an arbitrary reference frequency.
Applying our eccentricity detection statistic to \nsbh{}, we find that the eccentric hypothesis is favoured compared to the quasi-circular hypothesis with natural log Bayes factor $\ln {\cal B}\leq 0.9$, casting doubt on the eccentric interpretation for \nsbh.
  
\section{Eccentricity at different reference frequencies}
\label{sec:conversions}
The eccentricity of an inspiralling binary decreases with time due to gravitational-wave emission \citep{Peters1963, Peters64}.
Thus, when astronomers measure orbital eccentricity, it is typically defined at some \textit{reference frequency}. To further complicate matters, there are a number of different definitions of eccentricity in the literature, and the mapping between time and frequency depends on the choice of waveform approximant \citep{knee22}.
In this section, we investigate how priors and posteriors for eccentricity change depending on the choice of reference frequency. 
We show that the shape of these distributions varies strongly with the reference frequency. 
For example, a uniform prior defined at $\unit[20]{Hz}$ implies a very \textit{non-uniform} prior at $\unit[10]{Hz}$. 
This highlights the subtleties involved with trying to compare eccentricity measurements obtained at different reference frequencies and motivates the need for a less arbitrary method of characterising the presence of eccentricity. 

There are a number of tools available to evolve eccentricity from one frequency to another: 
\begin{itemize}
    \item \textit{Analytic approximations}: At leading order the eccentricity of a binary as a function of frequency is described in Ref.~\cite{Peters64}. Post-Newtonian corrections have been developed to improve the accuracy of the eccentricity evolution \citep[e.g.,][]{Konigsdorffer2006, Moore_2016, Fumagalli2025}. These formalisms are not waveform-dependent, so they predict the same eccentricity evolution for a binary system regardless of waveform used for the analysis. 
    \item \textit{Direct measurement}: The eccentricity evolution of a system can be directly measured from the gravitational waveform.~\texttt{gw-eccentricity} \citep{Shaikh2023, Shaikh2025} and \texttt{gwModels} \citep{Islam2025} are two python packages that allow the user to measure the eccentricity of a waveform using their own definitions of eccentricity and internal assumptions. 
\end{itemize}

\begin{figure}
    \centering
    \includegraphics[width=\linewidth]{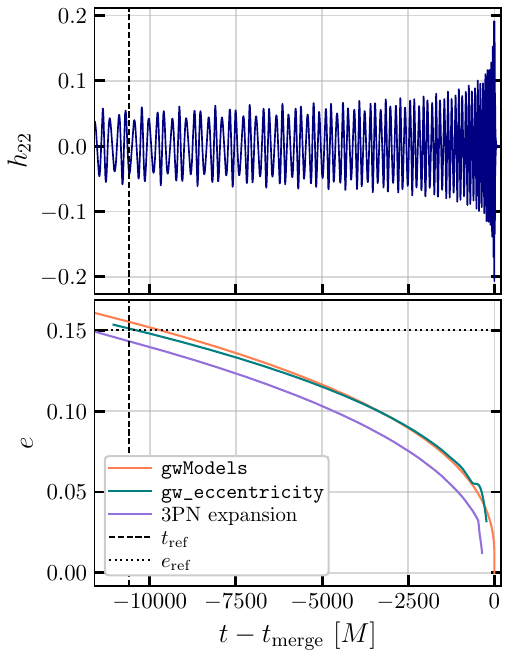}
    \caption{Top panel: The gravitational-waveform for a simulated merger with mass ratio $q=m_2/m_1 = 0.2$, generated using the \seob waveform model. We generate the waveform with an eccentricity of 0.15 at $\unit[20]{Hz}$ according to the \seob definition. Bottom panel: The eccentricity evolution for the waveform evolved with three different methods: PN expansion, \texttt{gw-eccentricity} and \texttt{gwModels}. We highlight the reference time (where $f=\unit[20]{Hz}$ according to \texttt{SEOBNRv5EHM}) with a  dashed vertical line.
    The initial eccentricity of $e=0.15$ is marked with the dotted horizontal line.}
    \label{fig:ecc_evolved}
\end{figure}

Figure~\ref{fig:ecc_evolved} shows the eccentricity evolution for a gravitational-wave signal with mass ratio $q=m_2/m_1=0.2$, similar to \nsbh{}, but with inflated masses for visualisation purposes (total mass $M=\unit[45]{M_\odot}$), with an initial eccentricity of $0.15$ at $\unit[20]{Hz}$, assuming the \texttt{SEOBNRv5EHM} definition of eccentricity. Each prescription results in a slightly different evolution due to differences in eccentricity definitions and methods of measuring the eccentricity, causing only one of the curves to actually cross $e=0.15$ at the dashed vertical line marking the reference time $t_\text{ref}$. At the reference time shown, where $e_{20}=\unit[0.15]{Hz}$, \gwecc and \texttt{gwModels} record eccentricities of 0.151 and 0.155, a difference of 0.8 and 3.4 percent from the \seob definition respectively.

\begin{figure}
\centering
\includegraphics[width=\columnwidth]{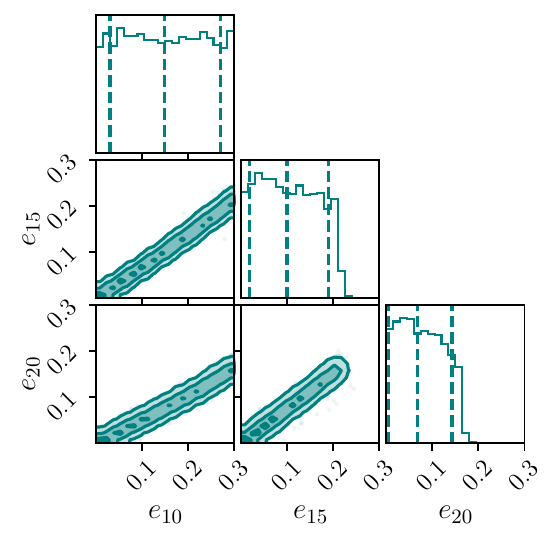}
    \caption{Prior samples drawn from a prior distribution with a reference frequency of $\unit[10]{Hz}$. The rows labelled  ``$e_{15}$'' and ``$e_{20}$'' represent the evolved prior to $\unit[15]{Hz}$ and $\unit[20]{Hz}$. A uniform eccentricity prior at $\unit[10]{Hz}$ is gradually transformed to a truncated, non-uniform distribution at $\unit[20]{Hz}$. }
    \label{fig:evolved_prior}
\end{figure}

To demonstrate the difference in priors between different reference frequencies, we sample from a standard set of priors for a binary black hole, including a uniform prior in eccentricity at $\unit[10]{Hz}$. 
For each sample, we generate the corresponding gravitational waveform using the \seob waveform.
We use \gwecc to evolve each eccentricity at $\unit[10]{Hz}$ to new eccentricities at $\unit[15]{Hz}$ and $\unit[20]{Hz}$.\footnote{We use the \texttt{ResidualAmplitude} method within \gwecc to measure the eccentricity. This is one of the recommended methods in Ref.~\cite{Shaikh2023}.}
Figure~\ref{fig:evolved_prior} shows the resulting eccentricity distributions and the correlations with each other. 
While the distribution of $e_{10}$ is uniform by definition, the distribution of $e_{20}$ is not. 

\section{Inference for \nsbh}
\label{sec:inference}
\subsection{Priors}
\label{sec:priors}
Given that priors can shift around depending on the choice of reference frequency, we argue that it is difficult to justify a particular choice of prior to perform inference on gravitational-wave events like \nsbh. 
In this section, we perform inference on \nsbh~using three commonly-employed prior distributions. Later we will show how these results can be re-processed with a marginalised eccentricity statistic.

The priors we choose to investigate (defined at a reference frequency of $\unit[20]{Hz}$) are: 
\begin{enumerate}
    \item Uniform in $e$ between 0 and 0.2. Many authors adopt a uniform in $e$ prior because the uniform distribution seems uninformative \citep[see, e.g.,][]{Morras2025}.
    \item Log-uniform in $e$ between $e=10^{-4}$ and $e=0.2$. Some authors have argued that this distribution is more suitable than a uniform distribution since we do not know the order of magnitude of eccentricity; see, e.g., Ref.~\citep{Lower18}. Arguably, this choice is more representative of at least some astrophysical distributions, where structure emerges when viewed on a log scale \citep[e.g.,][]{Zevin:2021:seleccentricity}.
    \item A physically-motivated distribution derived from simulating NSBH mergers in triples \citep{Stegmann2025}, evolved to $\unit[20]{Hz}$ and truncated between $e=0$ and $e=0.2$. 
\end{enumerate}

We perform an additional analysis at a reference frequency of $\unit[10]{Hz}$ using the NSBH triples prior evolved to $\unit[10]{Hz}$ and truncated between $e=0$ and $e=0.2$. 
We choose this relatively conservative upper limit on eccentricity to improve the computational efficiency of our inference.
In the next section we measure the eccentricity in \nsbh{} using each of these priors on eccentricity to demonstrate how the results vary with the choice of prior.

\subsection{Results}
\label{sec:results}

We use the spin-aligned, eccentric waveform model \texttt{SEOBNRv5EHM} \citep{Gamboa2024} with higher order multipoles to construct eccentricity posteriors for \nsbh. 
We use the Bayesian inference library \texttt{Bilby} \citep{bilby, Romero-Shaw:2020:Bilby} and \texttt{parallel\_bilby} \citep{Smith2020} to perform parameter estimation on the publicly available gravitational-wave strain data for \nsbh~ accessed via the Gravitational Wave Open Science Center (GWOSC)\citep{Abbott2023}.\footnote{Data products associated with the events analysed here are available through GWOSC at gw-openscience.org. The strain data for \nsbh~is available at \href{https://gwosc.org/eventapi/html/GWTC-3-marginal/GW200105_162426/v2/}{https://gwosc.org/eventapi/html/GWTC-3-marginal/GW200105\_162426/v2/}}

We analyse $\unit[32]{s}$ of data in LIGO Livingston and Virgo. 
We sample in eccentricity with the three priors described in Section~\ref{sec:priors}, with a starting (reference) orbit-averaged frequency of $\unit[20]{Hz}$ ($e_{20}$). 
We analyse $\unit[128]{s}$ of data for our additional test at a starting (reference) orbit-averaged frequency of $\unit[10]{Hz}$ ($e_{10}$). 
Although we sample uniformly in the relativistic anomaly (the second parameter necessary to completely characterise an eccentric binary), we expect the impact of this parameter on our inference to be small given the network signal-to-noise ratio $\text{SNR}\approx 14$ of \nsbh~\citep[e.g.,][]{clarke_2022}.\footnote{We note that for truly unbiased inference one should sample uniformly in mean anomaly rather than relativistic anomaly, since binaries spend more time close to apoapsis, but we expect this effect to be small.}
We sample from standard priors \citep[e.g.,][]{GWTC4_methods} in the binary parameters: uniform in the chirp mass, mass ratio, aligned spin components and  uniform in comoving volume and source frame time for luminosity distance.
We use the \texttt{dynesty} nested sampler \citep{dynesty} to sample with 1000 live points, distance and time marginalisation turned on, and a stopping criterion of $\Delta \text{log} \mathcal{Z} \leq 0.1$, where $\mathcal{Z}$ is the Bayesian evidence.

\begin{figure}
    \centering
    \includegraphics[width=\linewidth]{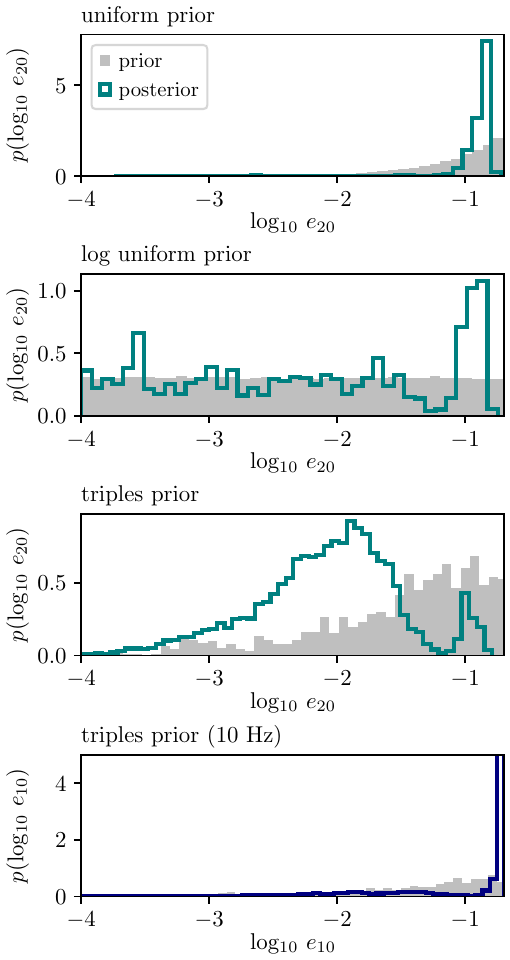}
    \caption{Posterior probability distributions for eccentricity for \nsbh~obtained using priors on eccentricity that are uniform, (top), log uniform (second panel), and informed by a distribution of NSBH mergers formed in a simulation of triples (third panel). The grey shaded regions show the prior distributions in each case. These results were obtained at a reference frequency of $\unit[20]{Hz}$. The bottom panel shows the posterior distribution for eccentricity obtained using the triple NSBH distribution at $\unit[10]{Hz}$. 
    At $\unit[20]{Hz}$, only the result obtained using a uniform prior on eccentricity (top panel) is peaked at high eccentricity, consistent with Ref.~\cite{Morras2025}. The $\unit[10]{Hz}$ result is also consistent with Ref.~\cite{Morras2025} if one evolves their result to $\unit[10]{Hz}$. 
    }
    \label{fig:GW200105}
\end{figure}

The results of our inference for the eccentricity of \nsbh\ are shown in Fig.~\ref{fig:GW200105} on a log-uniform eccentricity scale. 
Our uniform-prior results (second panel) are consistent with those from \cite{Morras2025, Planas2025, Kacanja2025, Jan2026}.
However, we find that at $\unit[20]{Hz}$, the uniform prior is the only one of the three analyses to peak confidently away from $e_{20}\approx 0$.
For the analysis at $\unit[10]{Hz}$ with $\unit[128]{s}$ of data and the NSBH triples prior (bottom panel), the posterior peaks at the edge of the prior boundary, $e_{10}=0.2$, which is consistent with the evolution of the uniform at $\unit[20]{Hz}$ posterior to $\unit[10]{Hz}$, but is less consistent with the evolution of the other $\unit[20]{Hz}$ analyses. This analysis also features strong prior support at high eccentricities. 

In addition, we analyse \nsbh~with a quasi-circular spin-precessing waveform model \texttt{SEOBNRv5PHM} \citep{Ramos-Buades2023} and with \seob with the eccentricity fixed to zero.
Taking these results at face value, we obtain natural log Bayes factors in support of the eccentric hypothesis over the quasi-circular precessing hypothesis of 2.4, 0.2 and 0.3 using a uniform, log-uniform and analytic NSBH triples prior respectively. 
We interpret this as weak, albeit prior-dependent, support in favour of eccentricity for \nsbh. 
However, the prior dependence for both the shape of the posterior on eccentricity and the Bayes factor motivate additional analyses. 
These results suggest that the apparent eccentricity in the signal, as observed in Refs.~\cite{Morras2025, Planas2025} and other studies, is ambiguous; results vary widely under multiple seemingly-reasonable choices of prior, and are also sensitive to the chosen reference frequency.
This raises the question of whether the eccentricity signal from Refs.~\cite{Morras2025,Planas2025} is significant, and if so, how one would know.

\section{Eccentricity detection statistic for \nsbh}
\label{sec:detection_stat}

\begin{figure}
    \centering
    \includegraphics[width=\linewidth]{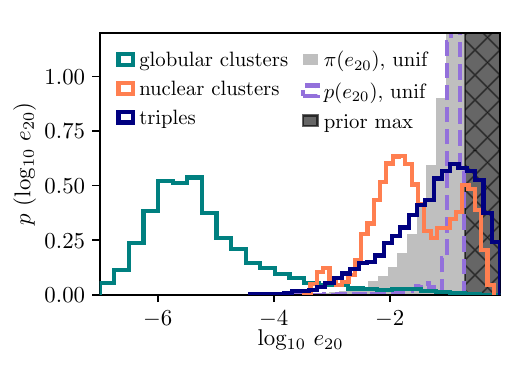}
    \caption{The three priors corresponding to dynamical formation channels we use to test the eccentric hypothesis for \nsbh\ are shown in solid histograms. The shaded region denotes values of eccentricity outside of the prior range of our inference (0.2 at $\unit[20]{Hz}$). For comparison, we show the shape of a uniform in $e$ prior at $\unit[20]{Hz}$ with the black dotted histogram, and the shape of the posterior on $e$ obtained when using this prior with the purple dashed histogram, similarly to Refs.~\citep[e.g.,][]{Morras2025, Planas2025, Jan2026}.}
    \label{fig:three_priors}
\end{figure}

Given the ambiguous results of the previous section, we propose an eccentricity detection statistic that marginalises over three astrophysically-motivated priors corresponding to formation channels that may produce eccentric NSBH mergers.
As we have demonstrated, eccentricity inferences can be extremely sensitive to prior choices. Even for low-mass NSBH mergers like \nsbh, the minimum measurable eccentricity at 10 Hz is $e_{10} \approx 0.03$ ($e_{20} \approx 0.015$) \citep{RomeroShaw2026}. Below this, the posterior will be heavily prior-dominated, so it is important to choose astrophysically-motivated priors.\footnote{We note that all dynamical channels in which gravitational-wave capture drives high-eccentricity mergers should converge to the same eccentricity distribution above some critical eccentricity \citep{Rozner2026}.} We show these distributions in Figure~\ref{fig:three_priors}: 

\begin{itemize}
    \item Globular clusters (GC). We use the globular cluster distribution of NSBH mergers from Refs.~\cite{ArcaSedda:2020:clusters, Dhurkunde2025}, available at \cite{dhurkunde_2023_data}. For NSBH mergers, globular clusters are not expected to produce many eccentric signals, because the majority of systems are expected to be ejected from the cluster and subsequently merge with a longer delay time that allows the residual eccentricity to radiate away before reaching the LVK band. Additionally, mass segregation in clusters tends to disfavour the pairing of neutron stars with black holes, making this an inefficient channel for NSBH mergers.
    
    \item Nuclear star clusters (NC). We use the distributions of NSBH mergers in nuclear star clusters that surround supermassive black holes in galactic centers from Refs.~\citep{Fragione2019, Fragione2019b, Dhurkunde2025}, available at \cite{dhurkunde_2023_data}. In these simulations, 10-20\% of the binaries enter the LVK observing band with eccentricities $\geq 0.1$, driven by von Zeipel-Kozai-Lidov resonances with the supermassive black hole or an additional stellar mass black hole within the cluster. 
    
    \item Triples (Tr). We use the NSBH distribution produced in the hierarchical field triples simulations from Ref.~\cite{Stegmann2025}, described in Section~\ref{sec:priors}. 
\end{itemize}

In principle, one should weigh the distributions by the relative rates of each channel, but without knowledge of the true formation channels of NSBH mergers or what the rates are, we assume they are the same. By doing so, we give enough support across the parameter space to enable downstream population studies to investigate the relative weight of each channel.
From here, we can define an effective prior given by:
\begin{equation}
   \pi_{\text{eff}}(e) = \frac{\pi_\text{GC}(e) + \pi_\text{NC}(e) + \pi_\text{Tr}(e)}{3}.
\end{equation}
The Bayesian evidence for this effective prior is given by:
\begin{equation}\label{eq:Z}
    \mathcal{Z}_\text{eff}(d) = \int de \ \mathcal{L}(d|e) \, \pi_\text{eff}(e).
\end{equation}
Importantly,  this evidence does not depend on the reference frequency, provided that the astrophysical distributions are evolved to the correct frequency.
We can evaluate Eq.~\ref{eq:Z} using a sum over posterior samples obtained with the fiducial prior (the prior for eccentricity used in the original inference) $\pi_{\o}(e)$; see, e.g., Ref.~\cite{ThraneTalbot18}:
\begin{equation}
    {\cal Z}_\text{eff}(d) = 
    \frac{{\cal Z}_{\o}(d)}{N}\sum^N_{i=1} \frac{\pi_\text{eff}(e_i) }
    {\pi_{\o}(e_i)} .
\end{equation}
Here, ${\cal Z}_{\o}(d)$ is the evidence obtained from the fiducial prior. While we opt to use a post-processing method to calculate $\mathcal{Z_\text{eff}}$, one could equivalently perform full parameter estimation with $\pi_\text{eff}$ or with the three astrophysical priors separately and combine with the desired weighting. 

We use this ${\cal Z}_\text{eff}$ to calculate two Bayes factors, one in favour of the eccentric hypothesis vs the $e=0$ hypothesis (both assuming aligned-spins), and another in favour of the precessing vs the aligned-spin hypothesis (as precession and eccentricity can be confused in certain regions of the parameter space \citep[e.g.,][]{Romero-Shaw:2023:EccOrPrecc, Tibrewal2026}). We compare the evidences calculated by SEOBNRv5EHM and by SEOBNRv5EHM with eccentricity fixed to 0 for the former, and we compare the evidences calculated by SEOBNRv5EHM with SEOBNRv5PHM for the latter. 
We find comparable natural log Bayes factors of $\ln {\cal B} \leq 0.4$ in the $e=0$ aligned spin case and $\ln {\cal B} \leq 0.9$ for the precessing quasi-circular case. 
Furthermore, we see a consistent decrease in the evidence when we test the detection statistic on a analysis performed at a reference frequency of $\unit[10]{Hz}$, with the NSBH triples prior evolved to $\unit[10]{Hz}$. This is the case despite the posterior peaking at higher eccentricity at $\unit[10]{Hz}$ and an increased length of data analysed in this test, since the support for eccentricity is partially driven by increased prior support at high eccentricities.
We summarise these results in Table~\ref{tab:BFs}. 

Of course, our statistic is only as reliable as the distributions contained within $\pi_\text{eff}$. 
Nonetheless, we believe our physically-motivated statistic offers a more defensible characterisation of the presence of eccentricity than any that depends on an arbitrary reference frequency.

\begin{table*}
    \centering
    \begin{tabular}{c|c|c|c|c}
    \label{tab:BFs}
     $\pi_\phi$,  f$_\text{ref}$    &  ln BF$_\text{ecc/prec}$, $\pi_\phi$  &  ln BF$_\text{ecc/prec}$, $\pi_\text{eff}$ & ln BF$_\text{ecc/e=0}$, $\pi_\phi$ & ln BF$_\text{ecc/e=0}$, $\pi_\text{eff}$\\
     \hline 
     \hline
     uniform $\unit[20]{Hz}$ & 2.4 & 0.9 & 1.9 & 0.4 \\
     log uniform $\unit[20]{Hz}$ & 0.2 & -0.5 & -0.2 & -0.9 \\
      triple $\unit[20]{Hz}$  &  0.3 &  -0.8 & -0.2 & -1.2 \\
      \hline 
      triple $\unit[10]{Hz}$ &  2.5 & 1.4 & -- & -- 
    \end{tabular}
    \caption{Summary of the natural log Bayes factors we obtain for the eccentric hypothesis. The first two columns record the natural log Bayes factor for the eccentric hypothesis vs the quasi-circular, precessing hypothesis for \nsbh\ for the fiducial prior used in parameter estimation and the astrophysically motivated prior $\pi_\text{eff}$. The last two columns show the same statistic obtained for eccentricity against the quasi-circular, aligned spin hypothesis. The final row shows the same statistic obtained from the NSBH triples prior evolved to $\unit[10]{Hz}$ and re-processed with $\pi_\text{eff}$ evolved to $\unit[10]{Hz}$. In this case the natural log Bayes factor is not directly comparable to the other entries in the table because different data is used ($\unit[128]{s}$ instead of $\unit[32]{s}$) but this result still follows the overall trend of decreased support for eccentricity. }
    \label{tab:placeholder}
\end{table*}

\section{Discussion and Conclusions}
\label{sec:discussion}
We conduct an investigation into possible eccentricity in the NSBH merger \nsbh{} \citep[e.g.,][]{Morras2025, Fei2024, Planas2025, Jan2026}.
We find that the inferred eccentricity is highly prior dependent (when using the same reference frequency), casting doubt on the eccentric hypothesis of \nsbh. 
 Moreover, we show that it is not straightforward to compare results obtained at different reference frequencies as they imply very different prior distributions when mapped to the same frequency. 
In the case of \nsbh, reweighting between e.g., a result obtained at $\unit[20]{Hz}$ to $\unit[10]{Hz}$ becomes prohibitively inefficient due to the large difference in data length when starting at these two frequencies. 

We replicate the results of Refs.~\cite{Morras2025, Planas2025, Jan2026, Kacanja2025} with the \texttt{SEOBNRv5EHM} waveform \citep{Gamboa2024} when we measure eccentricity with a uniform prior at $f_\text{ref} = \unit[20]{Hz}$.
However, we do not recover support for high eccentricities when using a log-uniform prior or a prior derived from NSBH triple simulations. 
Inspired by the prior dependence of our results, and lacking an obvious distribution in eccentricity to choose as a prior, we propose an \textit{eccentricity detection statistic} that marginalises over possible eccentricity distributions. 
Our demonstration decreases the support for non-zero eccentricity in \nsbh, although the accuracy of our result depends on how close to nature the distributions we include are, highlighting the importance of the research effort to understand the eccentricity distributions associated with the possible formation channels \citep[e.g.][]{Dorozsmai2026, Rozner2026}.

We caution that significance should be evaluated in the context of the population, rather than an event-by-event basis, due to trials factors, especially when interpreting the significance of potential outliers, like the eccentricity of \nsbh\ \citep{Mandel2026, Tenorio2026, Mould2026}.
There are currently very limited population analyses that include eccentricity \citep[see, e.g.,][]{Gupte:2024:eccentricity, Zeeshan2026, Morras2026}. Ref.~\cite{Morras2026} analyses the population of eccentric NSBH mergers, including \nsbh. With a small number of detections, the constraints on the eccentricity distribution are broad, but the inferred distribution is consistent with $\approx 1$ out of $3$ NSBH mergers being eccentric. 

With one potentially eccentric NSBH merger out of four detections with an inferred NSBH merger rate of $\unit[\approx 55]{Gpc^{-3}yr^{-1}}$ \citep{nsbh_detection, 230529, GWTC4, GWTC4_pop, Harry2026}, and assuming that, if eccentric, \nsbh\ came from a field triple, we estimate a rate of $\unit[\approx 15]{Gpc^{-3}yr^{-1}}$ of NSBH mergers from field triples (i.e., one quarter of the inferred rate of NSBH mergers). While the uncertainties on this rate are high, this is consistent with rates predicted for NSBH mergers in triple systems \citep{FragioneLoeb:2019:triples, HamersThompson:2019:triples} and could imply that most or all NSBH mergers are taking place in triples \citep{Stegmann2025, Stegmann2025b, RomeroShaw2026}. 
However, our results imply a lack of strong evidence for the triple formation channel when marginalising over dynamical formation channels. 

\section{Acknowledgements}

We thank Sylvia Biscoveanu, Snehal Tibrewal and Jacopo Tissino for thoughtful comments on the manuscript. 
This work is supported through Australian Research Council (ARC) Centres of Excellence CE170100004 and CE230900016, Discovery Projects DP220101610 and DP230103088, and LIEF Projects LE210100002 and LE260100008. 
T.\ A.\ C.\ acknowledges support from the Australian Government Research Training Program.
I.\ M.\ R-S gratefully acknowledges support from the Science and Technology Facilities Council Ernest Rutherford Fellowship grant number UKRI2423.
C.\ H. thanks the University of Portsmouth for support through the Dennis Sciama Fellowship.
This material is based upon work supported by NSF's LIGO Laboratory which is a major facility fully funded by the National Science Foundation.
The authors are grateful for for computational resources provided by the LIGO Laboratory computing cluster at California Institute of Technology supported by National Science Foundation Grants PHY-0757058 and PHY-0823459, the Ngarrgu Tindebeek / OzSTAR Australian national facility at Swinburne University of Technology and the Princeton Stellar computing cluster. We are also thankful for the computational resources provided by the DiRAC@Durham facility managed by the Institute for Computational Cosmology on behalf of the STFC DiRAC HPC Facility (www.dirac.ac.uk). The equipment was funded by BEIS capital funding via STFC capital grants ST/P002293/1, ST/R002371/1 and ST/S002502/1, Durham University and STFC operations grant ST/R000832/1. DiRAC is part of the National e-Infrastructure.

This research has made use of data or software obtained from the Gravitational Wave Open Science Center (gwosc.org), a service of the LIGO Scientific Collaboration, the Virgo Collaboration, and KAGRA. This material is based upon work supported by NSF's LIGO Laboratory which is a major facility fully funded by the National Science Foundation, as well as the Science and Technology Facilities Council (STFC) of the United Kingdom, the Max-Planck-Society (MPS), and the State of Niedersachsen/Germany for support of the construction of Advanced LIGO and construction and operation of the GEO600 detector. Additional support for Advanced LIGO was provided by the Australian Research Council. Virgo is funded, through the European Gravitational Observatory (EGO), by the French Centre National de Recherche Scientifique (CNRS), the Italian Istituto Nazionale di Fisica Nucleare (INFN) and the Dutch Nikhef, with contributions by institutions from Belgium, Germany, Greece, Hungary, Ireland, Japan, Monaco, Poland, Portugal, Spain. KAGRA is supported by Ministry of Education, Culture, Sports, Science and Technology (MEXT), Japan Society for the Promotion of Science (JSPS) in Japan; National Research Foundation (NRF) and Ministry of Science and ICT (MSIT) in Korea; Academia Sinica (AS) and National Science and Technology Council (NSTC) in Taiwan.

\section{Data Availability}
The data underlying this article will be shared on reasonable request to the corresponding author. This study made use of public gravitational-wave data sourced from GWOSC \citep{gwtc3,Abbott2023}. 

This work made use of the following software packages: \texttt{astropy} \citep{astropy:2013, astropy:2018, astropy:2022}, \texttt{matplotlib} \citep{Hunter:2007}, \texttt{numpy} \citep{numpy}, \texttt{python} \citep{python}, \texttt{scipy} \citep{2020SciPy-NMeth, scipy_11702230}, \texttt{Bilby} \citep{bilby, Bilby_2602178}, \texttt{dynesty} \citep{dynesty}, \gwecc \citep{Shaikh2023, Shaikh2025} and \texttt{corner.py} \citep{corner-Foreman-Mackey-2016, corner.py_4592454}.
Software citation information aggregated using \texttt{\href{https://www.tomwagg.com/software-citation-station/}{The Software Citation Station}} \citep{software-citation-station-paper, software-citation-station-zenodo}.

\bibliography{bib}

@String{mnras = "Mon. Not. R. Ast. Soc."}

@String{apj = "Astrophys. J."}

@String{apjl = "Astrophys. J. Lett."}

@String{prd = "Phys. Rev. D"}

@String{prl = "Phys. Rev. Lett."}

@String{pasa = "Pub. Astron. Soc. Aust."}

@ARTICLE{Naoz2013,
       author = {{Naoz}, Smadar and {Farr}, Will M. and {Lithwick}, Yoram and {Rasio}, Frederic A. and {Teyssandier}, Jean},
        title = "{Secular dynamics in hierarchical three-body systems}",
      journal = {Monthly Notices of the Royal Astronomical Society},
         year = 2013,
        month = may,
       volume = {431},
       number = {3},
        pages = {2155-2171},
          doi = {10.1093/mnras/stt302},
archivePrefix = {arXiv},
       eprint = {1107.2414},
 primaryClass = {astro-ph.EP},
       adsurl = {https://ui.adsabs.harvard.edu/abs/2013MNRAS.431.2155N}
}

@ARTICLE{Gayathri_2022,
       author = {{Gayathri}, V. and {Healy}, J. and {Lange}, J. and {O'Brien}, B. and {Szczepa{\'n}czyk}, M. and {Bartos}, Imre and {Campanelli}, M. and {Klimenko}, S. and {Lousto}, C.~O. and {O'Shaughnessy}, R.},
        title = "{Eccentricity estimate for black hole mergers with numerical relativity simulations}",
      journal = {Nature Astronomy},
         year = 2022,
        month = jan,
       volume = {6},
        pages = {344-349},
          doi = {10.1038/s41550-021-01568-w},
       adsurl = {https://ui.adsabs.harvard.edu/abs/2022NatAs...6..344G}
}

@article{Romero-Shaw:2019itr,
    author = "Romero-Shaw, Isobel M. and Lasky, Paul D. and Thrane, Eric",
    title = "{Searching for Eccentricity: Signatures of Dynamical Formation in the First Gravitational-Wave Transient Catalogue of LIGO and Virgo}",
    eprint = "1909.05466",
    archivePrefix = "arXiv",
    primaryClass = "astro-ph.HE",
    doi = "10.1093/mnras/stz2996",
    journal = "Mon. Not. Roy. Astron. Soc.",
    volume = "490",
    number = "4",
    pages = "5210--5216",
    year = "2019"
}

@ARTICLE{Romero-Shaw2021,
       author = {{Romero-Shaw}, Isobel and {Lasky}, Paul D. and {Thrane}, Eric},
        title = "{Signs of Eccentricity in Two Gravitational-wave Signals May Indicate a Subpopulation of Dynamically Assembled Binary Black Holes}",
      journal = {\apjl},
         year = 2021,
        month = nov,
       volume = {921},
       number = {2},
          eid = {L31},
        pages = {L31},
          doi = {10.3847/2041-8213/ac3138},
archivePrefix = {arXiv},
       eprint = {2108.01284},
 primaryClass = {astro-ph.HE},
       adsurl = {https://ui.adsabs.harvard.edu/abs/2021ApJ...921L..31R}
}

@article{Romero-Shaw:2020:GW190521,
    author = "Romero-Shaw, Isobel M. and Lasky, Paul D. and Thrane, Eric and Bustillo, Juan Calderon",
    title = "{GW190521: orbital eccentricity and signatures of dynamical formation in a binary black hole merger signal}",
    eprint = "2009.04771",
    archivePrefix = "arXiv",
    primaryClass = "astro-ph.HE",
    doi = "10.3847/2041-8213/abbe26",
    journal = "Astrophys. J. Lett.",
    volume = "903",
    number = "1",
    pages = "L5",
    year = "2020"
}

@ARTICLE{Gamba_2021,
       author = {{Gamba}, R. and {Breschi}, M. and {Carullo}, G. and {Albanesi}, S. and {Rettegno}, P. and {Bernuzzi}, S. and {Nagar}, A.},
        title = "{GW190521 as a dynamical capture of two nonspinning black holes}",
      journal = {Nature Astronomy},
         year = 2023,
        month = jan,
       volume = {7},
        pages = {11-17},
          doi = {10.1038/s41550-022-01813-w},
archivePrefix = {arXiv},
       eprint = {2106.05575},
 primaryClass = {gr-qc},
       adsurl = {https://ui.adsabs.harvard.edu/abs/2023NatAs...7...11G}
}

@article{Romero-Shaw:2020:Bilby,
    author = "Romero-Shaw, I. M. and others",
    title = "{Bayesian inference for compact binary coalescences with bilby: validation and application to the first LIGO\textendash{}Virgo gravitational-wave transient catalogue}",
    eprint = "2006.00714",
    archivePrefix = "arXiv",
    primaryClass = "astro-ph.IM",
    doi = "10.1093/mnras/staa2850",
    journal = "Mon. Not. Roy. Astron. Soc.",
    volume = "499",
    number = "3",
    pages = "3295--3319",
    year = "2020"
}

@ARTICLE{Zevin:2021:seleccentricity,
       author = {{Zevin}, Michael and {Romero-Shaw}, Isobel M. and {Kremer}, Kyle and {Thrane}, Eric and {Lasky}, Paul D.},
        title = "{Implications of Eccentric Observations on Binary Black Hole Formation Channels}",
      journal = {\apjl},
         year = 2021,
        month = nov,
       volume = {921},
       number = {2},
          eid = {L43},
        pages = {L43},
          doi = {10.3847/2041-8213/ac32dc},
archivePrefix = {arXiv},
       eprint = {2106.09042},
 primaryClass = {astro-ph.HE},
       adsurl = {https://ui.adsabs.harvard.edu/abs/2021ApJ...921L..43Z}
}

@article{2020_Kagra,
author = {Akutsu, T and Ando, M and Arai, K and Arai, Yasuo and Araki, Sakae and Araya, Akito and Aritomi, N and Aso, Yoichi and Bae, Susung and Bae, Y and Baiotti, Luca and Bajpai, Rishabh and Barton, M. and Cannon, Kipp and Capocasa, E and Chan, Mariodicristo and Chen, Chanratana and Chen, K and Chen, Y},
year = {2020},
month = {08},
pages = {},
title = "{Overview of KAGRA : Detector design and construction history}",
volume = {2021},
journal = {Progress of Theoretical and Experimental Physics},
doi = {10.1093/ptep/ptaa125}
}

@article{adv_ligo_2015,
   title="{Advanced LIGO}",
   volume={32},
   ISSN={1361-6382},
   url={http://dx.doi.org/10.1088/0264-9381/32/7/074001},
   DOI={10.1088/0264-9381/32/7/074001},
   number={7},
   journal={Classical and Quantum Gravity},
   publisher={IOP Publishing},
   author={Aasi, J and Abbott, B P and Abbott, R and Abbott, T and Abernathy, M R and Ackley, K and Adams, C and Adams, T and Addesso, P and et al.},
   year={2015},
   month={Mar},
   pages={074001}
}

@ARTICLE{Dhurkunde2025,
       author = {{Dhurkunde}, Rahul and {Nitz}, Alexander H.},
        title = "{Search for eccentric NSBH and BNS mergers in the third observing run of Advanced LIGO and Virgo}",
      journal = {Physical Review D},
         year = 2025,
        month = may,
       volume = {111},
       number = {10},
          eid = {103018},
        pages = {103018},
          doi = {10.1103/PhysRevD.111.103018},
archivePrefix = {arXiv},
       eprint = {2311.00242},
 primaryClass = {astro-ph.HE},
       adsurl = {https://ui.adsabs.harvard.edu/abs/2025PhRvD.111j3018D}
}

@ARTICLE{HamersThompson:2019:triples,
       author = {{Hamers}, Adrian S. and {Thompson}, Todd A.},
        title = "{Double Neutron Star Mergers from Hierarchical Triple-star Systems}",
      journal = {\apj},
         year = 2019,
        month = sep,
       volume = {883},
       number = {1},
          eid = {23},
        pages = {23},
          doi = {10.3847/1538-4357/ab3b06},
archivePrefix = {arXiv},
       eprint = {1907.08297},
 primaryClass = {astro-ph.HE},
       adsurl = {https://ui.adsabs.harvard.edu/abs/2019ApJ...883...23H}
}

@ARTICLE{FragioneLoeb:2019:triples,
       author = {{Fragione}, Giacomo and {Loeb}, Abraham},
        title = "{Black hole-neutron star mergers from triples}",
      journal = {\mnras},
         year = 2019,
        month = jul,
       volume = {486},
       number = {3},
        pages = {4443-4450},
          doi = {10.1093/mnras/stz1131},
archivePrefix = {arXiv},
       eprint = {1903.10511},
 primaryClass = {astro-ph.GA},
       adsurl = {https://ui.adsabs.harvard.edu/abs/2019MNRAS.486.4443F}
}

@ARTICLE{ArcaSedda:2020:clusters,
       author = {{Arca Sedda}, Manuel},
        title = "{Dissecting the properties of neutron star-black hole mergers originating in dense star clusters}",
      journal = {Communications Physics},
         year = 2020,
        month = mar,
       volume = {3},
       number = {1},
          eid = {43},
        pages = {43},
          doi = {10.1038/s42005-020-0310-x},
archivePrefix = {arXiv},
       eprint = {2003.02279},
 primaryClass = {astro-ph.GA},
       adsurl = {https://ui.adsabs.harvard.edu/abs/2020CmPhy...3...43A}
}

@ARTICLE{Romero-Shaw:2023:EccOrPrecc,
       author = {{Romero-Shaw}, Isobel M. and {Gerosa}, Davide and {Loutrel}, Nicholas},
        title = "{Eccentricity or spin precession? Distinguishing subdominant effects in gravitational-wave data}",
      journal = {\mnras},
         year = 2023,
        month = mar,
       volume = {519},
       number = {4},
        pages = {5352-5357},
          doi = {10.1093/mnras/stad031},
archivePrefix = {arXiv},
       eprint = {2211.07528},
 primaryClass = {astro-ph.HE},
       adsurl = {https://ui.adsabs.harvard.edu/abs/2023MNRAS.519.5352R}
}

@article{AdvancedVirgo,
      author         = "Acernese, F. and others",
      title          = "{Advanced Virgo: a second-generation interferometric
                        gravitational wave detector}",
      collaboration  = "VIRGO",
      journal        = "Class. Quant. Grav.",
      volume         = "32",
      year           = "2015",
      number         = "2",
      pages          = "024001",
      doi            = "10.1088/0264-9381/32/2/024001",
      eprint         = "1408.3978",
      archivePrefix  = "arXiv",
      primaryClass   = "gr-qc",
}

@ARTICLE{Gupte:2024:eccentricity,
       author = {{Gupte}, Nihar and {Ramos-Buades}, Antoni and {Buonanno}, Alessandra and {Gair}, Jonathan and {Coleman Miller}, M. and {Dax}, Maximilian and {Green}, Stephen R. and {P{\"u}rrer}, Michael and {Wildberger}, Jonas and {Macke}, Jakob and et al.},
        title = "{Evidence for eccentricity in the population of binary black holes observed by LIGO-Virgo-KAGRA}",
      journal = {Physical Review D},
         year = 2025,
        month = nov,
       volume = {112},
       number = {10},
          eid = {104045},
        pages = {104045},
          doi = {10.1103/vpyp-nvfp},
archivePrefix = {arXiv},
       eprint = {2404.14286},
 primaryClass = {gr-qc},
       adsurl = {https://ui.adsabs.harvard.edu/abs/2025PhRvD.112j4045G}
}

@ARTICLE{Ramos-Buades:2023:analysis,
       author = {{Ramos-Buades}, Antoni and {Buonanno}, Alessandra and {Gair}, Jonathan},
        title = "{Bayesian inference of binary black holes with inspiral-merger-ringdown waveforms using two eccentric parameters}",
      journal = {\prd},
         year = 2023,
        month = dec,
       volume = {108},
       number = {12},
          eid = {124063},
        pages = {124063},
          doi = {10.1103/PhysRevD.108.124063},
archivePrefix = {arXiv},
       eprint = {2309.15528},
 primaryClass = {gr-qc},
       adsurl = {https://ui.adsabs.harvard.edu/abs/2023PhRvD.108l4063R}
}

@ARTICLE{RodriguezAntonini:2018:triples,
       author = {{Rodriguez}, Carl L. and {Antonini}, Fabio},
        title = "{A Triple Origin for the Heavy and Low-spin Binary Black Holes Detected by LIGO/VIRGO}",
      journal = {\apj},
         year = 2018,
        month = aug,
       volume = {863},
       number = {1},
          eid = {7},
        pages = {7},
          doi = {10.3847/1538-4357/aacea4},
archivePrefix = {arXiv},
       eprint = {1805.08212},
 primaryClass = {astro-ph.HE},
       adsurl = {https://ui.adsabs.harvard.edu/abs/2018ApJ...863....7R}
}

@ARTICLE{vonZeipel1910,
       author = {{von Zeipel}, H.},
        title = "{Sur l'application des s{\'e}ries de M. Lindstedt {\`a} l'{\'e}tude du mouvement des com{\`e}tes p{\'e}riodiques}",
      journal = {Astronomische Nachrichten},
         year = 1910,
        month = mar,
       volume = {183},
       number = {22},
        pages = {345},
          doi = {10.1002/asna.19091832202},
       adsurl = {https://ui.adsabs.harvard.edu/abs/1910AN....183..345V}
}

@ARTICLE{Antonini:2025:hierarchical,
       author = {{Antonini}, Fabio and {Romero-Shaw}, Isobel M. and {Callister}, Thomas},
        title = "{Star Cluster Population of High Mass Black Hole Mergers in Gravitational Wave Data}",
      journal = {\prl},
         year = 2025,
        month = jan,
       volume = {134},
       number = {1},
          eid = {011401},
        pages = {011401},
          doi = {10.1103/PhysRevLett.134.011401},
archivePrefix = {arXiv},
       eprint = {2406.19044},
 primaryClass = {astro-ph.HE},
       adsurl = {https://ui.adsabs.harvard.edu/abs/2025PhRvL.134a1401A}
}

@ARTICLE{Oleary2009,
       author = {{O'Leary}, Ryan M. and {Kocsis}, Bence and {Loeb}, Abraham},
        title = "{Gravitational waves from scattering of stellar-mass black holes in galactic nuclei}",
      journal = {\mnras},
         year = 2009,
        month = jun,
       volume = {395},
       number = {4},
        pages = {2127-2146},
          doi = {10.1111/j.1365-2966.2009.14653.x},
archivePrefix = {arXiv},
       eprint = {0807.2638},
 primaryClass = {astro-ph},
       adsurl = {https://ui.adsabs.harvard.edu/abs/2009MNRAS.395.2127O}
}

@article{Rodriguez18b,
      author         = "Rodriguez, Carl L. and Amaro-Seoane, Pau and Chatterjee,
                        Sourav and Kremer, Kyle and Rasio, Frederic A. and
                        Samsing, Johan and Ye, Claire S. and Zevin, Michael",
      title          = "{Post-Newtonian Dynamics in Dense Star Clusters:
                        Formation, Masses, and Merger Rates of Highly-Eccentric
                        Black Hole Binaries}",
      journal        = "Phys. Rev.",
      volume         = "D98",
      year           = "2018",
      number         = "12",
      pages          = "123005",
      doi            = "10.1103/PhysRevD.98.123005",
      eprint         = "1811.04926",
      archivePrefix  = "arXiv",
      primaryClass   = "astro-ph.HE",
}

@ARTICLE{GWTC-2_RnP,
       author = {{Abbott}, R. and {Abbott}, T.~D. and {Abraham}, S. and {Acernese}, F. and
         {Ackley}, K. and {Adams}, A. and {Adams}, C. and others},
    title = "{Population Properties of Compact Objects from the Second LIGO-Virgo Gravitational-Wave Transient Catalog}",
      journal = {\apjl},
         year = 2021,
        month = may,
       volume = {913},
       number = {1},
          eid = {L7},
        pages = {L7},
          doi = {10.3847/2041-8213/abe949},
archivePrefix = {arXiv},
       eprint = {2010.14533},
 primaryClass = {astro-ph.HE},
       adsurl = {https://ui.adsabs.harvard.edu/abs/2021ApJ...913L...7A}
}

@ARTICLE{Kozai62,
   author = {{Kozai}, Y.},
    title = "{Secular perturbations of asteroids with high inclination and eccentricity}",
  journal = {Astrophys. J.},
     year = 1962,
    month = nov,
   volume = 67,
    pages = {591},
      doi = {10.1086/108790},
   adsurl = {http://adsabs.harvard.edu/abs/1962AJ.....67..591K}
}

@ARTICLE{Lidov62,
   author = {{Lidov}, M.~L.},
    title = "{The evolution of orbits of artificial satellites of planets under the action of gravitational perturbations of external bodies}",
  journal = {Planetary and Space Science},
     year = 1962,
    month = oct,
   volume = 9,
    pages = {719-759},
      doi = {10.1016/0032-0633(62)90129-0},
   adsurl = {http://adsabs.harvard.edu/abs/1962Po26SS....9..719L}
}

@article{Peters64,
  title = {Gravitational Radiation and the Motion of Two Point Masses},
  author = {Peters, P. C.},
  journal = {Phys. Rev.},
  volume = {136},
  issue = {4B},
  pages = {B1224--B1232},
  numpages = {0},
  year = {1964},
  month = {Nov},
  publisher = {American Physical Society},
  doi = {10.1103/PhysRev.136.B1224},
}

@ARTICLE{Peters1963,
       author = {{Peters}, P.~C. and {Mathews}, J.},
        title = "{Gravitational Radiation from Point Masses in a Keplerian Orbit}",
      journal = {Physical Review},
         year = 1963,
        month = jul,
       volume = {131},
       number = {1},
        pages = {435-440},
          doi = {10.1103/PhysRev.131.435},
       adsurl = {https://ui.adsabs.harvard.edu/abs/1963PhRv..131..435P}
}

@ARTICLE{Abac2025,
       author = {{Abac}, A.~G. and {Abouelfettouh}, I. and {Acernese}, F. and {Ackley}, K. and {Adamcewicz}, C. and {Adhicary}, S. and {Adhikari}, D. and {Adhikari}, N. and {Adhikari}, R.~X. and {Adkins}, V.~K. and et al.},
        title = "{GW241011 and GW241110: Exploring Binary Formation and Fundamental Physics with Asymmetric, High-spin Black Hole Coalescences}",
      journal = {The Astrophysical Journal},
         year = 2025,
        month = nov,
       volume = {993},
       number = {1},
          eid = {L21},
        pages = {L21},
          doi = {10.3847/2041-8213/ae0d54},
archivePrefix = {arXiv},
       eprint = {2510.26931},
 primaryClass = {astro-ph.HE},
       adsurl = {https://ui.adsabs.harvard.edu/abs/2025ApJ...993L..21A}
}

@ARTICLE{GWTC4_methods,
       author = {{The LIGO Scientific Collaboration} and {the Virgo Collaboration} and {the KAGRA Collaboration} and {Abac}, A.~G. and {Abouelfettouh}, I. and {Acernese}, F. and {Ackley}, K. and {Adhicary}, S. and {Adhikari}, D. and {Adhikari}, N. and et al.},
        title = "{GWTC-4.0: Methods for Identifying and Characterizing Gravitational-wave Transients}",
      journal = {arXiv e-prints},
         year = 2025,
        month = aug,
          eid = {arXiv:2508.18081},
        pages = {arXiv:2508.18081},
          doi = {10.48550/arXiv.2508.18081},
archivePrefix = {arXiv},
       eprint = {2508.18081},
 primaryClass = {gr-qc},
       adsurl = {https://ui.adsabs.harvard.edu/abs/2025arXiv250818081T}
}

@ARTICLE{Lower18,
       author = {{Lower}, Marcus E. and {Thrane}, Eric and {Lasky}, Paul D. and {Smith}, Rory},
        title = "{Measuring eccentricity in binary black hole inspirals with gravitational waves}",
      journal = {\prd},
         year = 2018,
        month = oct,
       volume = {98},
       number = {8},
          eid = {083028},
        pages = {083028},
          doi = {10.1103/PhysRevD.98.083028},
archivePrefix = {arXiv},
       eprint = {1806.05350},
 primaryClass = {astro-ph.HE},
       adsurl = {https://ui.adsabs.harvard.edu/abs/2018PhRvD..98h3028L}
}

@ARTICLE{Wu_2020,
       author = {{Wu}, Shichao and {Cao}, Zhoujian and {Zhu}, Zong-Hong},
        title = "{Measuring the eccentricity of binary black holes in GWTC-1 by using the inspiral-only waveform}",
      journal = {\mnras},
         year = 2020,
        month = jun,
       volume = {495},
       number = {1},
        pages = {466-478},
          doi = {10.1093/mnras/staa1176},
archivePrefix = {arXiv},
       eprint = {2002.05528},
 primaryClass = {astro-ph.IM},
       adsurl = {https://ui.adsabs.harvard.edu/abs/2020MNRAS.495..466W}
}

@article{Naoz_2016,
   title="{The Eccentric Kozai-Lidov Effect and Its Applications}",
   volume={54},
   ISSN={1545-4282},
   url={http://dx.doi.org/10.1146/annurev-astro-081915-023315},
   DOI={10.1146/annurev-astro-081915-023315},
   number={1},
   journal={Annual Review of Astronomy and Astrophysics},
   publisher={Annual Reviews},
   author={Naoz, Smadar},
   year={2016},
   month={Sep},
   pages={441–489}
}

@ARTICLE{OSheaKumar2021,
       author = {{O'Shea}, Eamonn and {Kumar}, Prayush},
        title = "{Correlations in gravitational-wave reconstructions from eccentric binaries: A case study with GW151226 and GW170608}",
      journal = {\prd},
         year = 2023,
        month = nov,
       volume = {108},
       number = {10},
          eid = {104018},
        pages = {104018},
          doi = {10.1103/PhysRevD.108.104018},
archivePrefix = {arXiv},
       eprint = {2107.07981},
 primaryClass = {astro-ph.HE},
       adsurl = {https://ui.adsabs.harvard.edu/abs/2023PhRvD.108j4018O}
}

@ARTICLE{dynesty,
       author = {{Speagle}, Joshua S.},
        title = "{DYNESTY: a dynamic nested sampling package for estimating Bayesian posteriors and evidences}",
      journal = {Monthly Notices of the Royal Astronomical Society},
         year = 2020,
        month = apr,
       volume = {493},
       number = {3},
        pages = {3132-3158},
          doi = {10.1093/mnras/staa278},
archivePrefix = {arXiv},
       eprint = {1904.02180},
 primaryClass = {astro-ph.IM},
       adsurl = {https://ui.adsabs.harvard.edu/abs/2020MNRAS.493.3132S}
}

@article{bilby,
      author         = "Ashton, Gregory and others",
      title          = "{Bilby: A user-friendly Bayesian inference library for
                        gravitational-wave astronomy}",
      journal        = "Astrophys. J. Suppl.",
      volume         = "241",
      year           = "2019",
      number         = "2",
      pages          = "27",
      doi            = "10.3847/1538-4365/ab06fc",
      eprint         = "1811.02042",
      archivePrefix  = "arXiv",
      primaryClass   = "astro-ph.IM",
}

@ARTICLE{ThraneTalbot18,
       author = {{Thrane}, Eric and {Talbot}, Colm},
        title = "{An introduction to Bayesian inference in gravitational-wave astronomy: Parameter estimation, model selection, and hierarchical models}",
      journal = {\pasa},
         year = 2019,
        month = mar,
       volume = {36},
          eid = {e010},
        pages = {e010},
          doi = {10.1017/pasa.2019.2},
archivePrefix = {arXiv},
       eprint = {1809.02293},
 primaryClass = {astro-ph.IM},
       adsurl = {https://ui.adsabs.harvard.edu/abs/2019PASA...36...10T}
}

@ARTICLE{Petrovich2017,
       author = {{Petrovich}, Cristobal and {Antonini}, Fabio},
        title = "{Greatly Enhanced Merger Rates of Compact-object Binaries in Non-spherical Nuclear Star Clusters}",
      journal = {The Astrophysical Journal},
         year = 2017,
        month = sep,
       volume = {846},
       number = {2},
          eid = {146},
        pages = {146},
          doi = {10.3847/1538-4357/aa8628},
archivePrefix = {arXiv},
       eprint = {1705.05848},
 primaryClass = {astro-ph.HE},
       adsurl = {https://ui.adsabs.harvard.edu/abs/2017ApJ...846..146P}
}

@ARTICLE{Ye2020,
       author = {{Ye}, Claire S. and {Fong}, Wen-fai and {Kremer}, Kyle and {Rodriguez}, Carl L. and {Chatterjee}, Sourav and {Fragione}, Giacomo and {Rasio}, Frederic A.},
        title = "{On the Rate of Neutron Star Binary Mergers from Globular Clusters}",
      journal = {The Astrophysical Journal},
         year = 2020,
        month = jan,
       volume = {888},
       number = {1},
          eid = {L10},
        pages = {L10},
          doi = {10.3847/2041-8213/ab5dc5},
archivePrefix = {arXiv},
       eprint = {1910.10740},
 primaryClass = {astro-ph.HE},
       adsurl = {https://ui.adsabs.harvard.edu/abs/2020ApJ...888L..10Y}
}

@ARTICLE{Fragione2020,
       author = {{Fragione}, Giacomo and {Banerjee}, Sambaran},
        title = "{Demographics of Neutron Stars in Young Massive and Open Clusters}",
      journal = {The Astrophysical Journal},
         year = 2020,
        month = sep,
       volume = {901},
       number = {1},
          eid = {L16},
        pages = {L16},
          doi = {10.3847/2041-8213/abb671},
archivePrefix = {arXiv},
       eprint = {2006.06702},
 primaryClass = {astro-ph.GA},
       adsurl = {https://ui.adsabs.harvard.edu/abs/2020ApJ...901L..16F}
}

@ARTICLE{Mandel2026,
       author = {{Mandel}, Ilya},
        title = "{What is the Most Massive Gravitational-wave Source?}",
      journal = {The Astrophysical Journal},
         year = 2026,
        month = jan,
       volume = {996},
       number = {1},
          eid = {L4},
        pages = {L4},
          doi = {10.3847/2041-8213/ae278d},
archivePrefix = {arXiv},
       eprint = {2509.05885},
 primaryClass = {astro-ph.HE},
       adsurl = {https://ui.adsabs.harvard.edu/abs/2026ApJ...996L...4M}
}

@ARTICLE{Tenorio2026,
       author = {{Tenorio}, Rodrigo and {Gerosa}, Davide},
        title = "{On the exceptionality of exceptional gravitational-wave events}",
      journal = {arXiv e-prints},
         year = 2026,
        month = jan,
          eid = {arXiv:2601.02467},
        pages = {arXiv:2601.02467},
          doi = {10.48550/arXiv.2601.02467},
archivePrefix = {arXiv},
       eprint = {2601.02467},
 primaryClass = {astro-ph.HE},
       adsurl = {https://ui.adsabs.harvard.edu/abs/2026arXiv260102467T}
}

@ARTICLE{Mould2026,
       author = {{Mould}, Matthew and {Tenorio}, Rodrigo and {Gerosa}, Davide},
        title = "{Gravitational-wave astronomy requires population-informed parameter estimation}",
      journal = {arXiv e-prints},
         year = 2026,
        month = apr,
          eid = {arXiv:2604.15885},
        pages = {arXiv:2604.15885},
          doi = {10.48550/arXiv.2604.15885},
archivePrefix = {arXiv},
       eprint = {2604.15885},
 primaryClass = {gr-qc},
       adsurl = {https://ui.adsabs.harvard.edu/abs/2026arXiv260415885M}
}

@ARTICLE{Morras2026,
       author = {{Morras}, Gonzalo and {Pratten}, Geraint and {Schmidt}, Patricia},
        title = "{Impact of eccentricity on the population properties of neutron star - black hole mergers}",
      journal = {arXiv e-prints},
         year = 2026,
        month = mar,
          eid = {arXiv:2603.22461},
        pages = {arXiv:2603.22461},
          doi = {10.48550/arXiv.2603.22461},
archivePrefix = {arXiv},
       eprint = {2603.22461},
 primaryClass = {astro-ph.HE},
       adsurl = {https://ui.adsabs.harvard.edu/abs/2026arXiv260322461M}
}

@ARTICLE{Hannam2022,
       author = {{Hannam}, Mark and {Hoy}, Charlie and {Thompson}, Jonathan E. and {Fairhurst}, Stephen and {Raymond}, Vivien and {Colleoni}, Marta and {Davis}, Derek and {Estell{\'e}s}, H{\'e}ctor and {Haster}, Carl-Johan and {Helmling-Cornell}, Adrian and et al.},
        title = "{General-relativistic precession in a black-hole binary}",
      journal = {Nature},
         year = 2022,
        month = oct,
       volume = {610},
       number = {7933},
        pages = {652-655},
          doi = {10.1038/s41586-022-05212-z},
archivePrefix = {arXiv},
       eprint = {2112.11300},
 primaryClass = {gr-qc},
       adsurl = {https://ui.adsabs.harvard.edu/abs/2022Natur.610..652H}
}

@ARTICLE{Harry2026,
       author = {{Harry}, Ian and {Hoy}, Charlie},
        title = "{Constraining the neutron star─black hole merger rate}",
      journal = {Physical Review D},
         year = 2026,
        month = jan,
       volume = {113},
       number = {2},
          eid = {L021305},
        pages = {L021305},
          doi = {10.1103/cqqn-gl4y},
archivePrefix = {arXiv},
       eprint = {2503.09773},
 primaryClass = {hep-ex},
       adsurl = {https://ui.adsabs.harvard.edu/abs/2026PhRvD.113b1305H}
}

@ARTICLE{Apostolatos1994,
       author = {{Apostolatos}, Theocharis A. and {Cutler}, Curt and {Sussman}, Gerald J. and {Thorne}, Kip S.},
        title = "{Spin-induced orbital precession and its modulation of the gravitational waveforms from merging binaries}",
      journal = {Physical Review D},
         year = 1994,
        month = jun,
       volume = {49},
       number = {12},
        pages = {6274-6297},
          doi = {10.1103/PhysRevD.49.6274},
       adsurl = {https://ui.adsabs.harvard.edu/abs/1994PhRvD..49.6274A}
}

@ARTICLE{Zeeshan2026,
       author = {{Zeeshan}, Muhammad and {O'Shaughnessy}, Richard and {Malagon}, Natalie},
        title = "{Population Properties of Binary Black Holes with Eccentricity}",
      journal = {arXiv e-prints},
         year = 2026,
        month = feb,
          eid = {arXiv:2602.11030},
        pages = {arXiv:2602.11030},
          doi = {10.48550/arXiv.2602.11030},
archivePrefix = {arXiv},
       eprint = {2602.11030},
 primaryClass = {astro-ph.HE},
       adsurl = {https://ui.adsabs.harvard.edu/abs/2026arXiv260211030Z}
}

@article{Moore_2016,
   title={Gravitational-wave phasing for low-eccentricity inspiralling compact binaries to 3PN order},
   volume={93},
   ISSN={2470-0029},
   url={http://dx.doi.org/10.1103/PhysRevD.93.124061},
   DOI={10.1103/physrevd.93.124061},
   number={12},
   journal={Physical Review D},
   publisher={American Physical Society (APS)},
   author={Moore, Blake and Favata, Marc and Arun, K. G. and Mishra, Chandra Kant},
   year={2016},
   month={Jun}
}

@article{nsbh_detection,
	doi = {10.3847/2041-8213/ac082e},
	url = {https://doi.org/10.3847/2041-8213/ac082e},
	year = 2021,
	month = {jun},
	publisher = {American Astronomical Society},
	volume = {915},
	number = {1},
	pages = {L5},
	author = {R. Abbott and others},
	title = "{Observation of Gravitational Waves from Two Neutron Star{\textendash}Black Hole Coalescences}",
	journal = {\apjl}
	}

@ARTICLE{gwtc3,
       author = {{Abbott}, R. and {Abbott}, T.~D. and {Acernese}, F. and {Ackley}, K. and {Adams}, C. and {Adhikari}, N. and {Adhikari}, R.~X. and {Adya}, V.~B. and {Affeldt}, C. and {Agarwal}, D. and et al.},
        title = "{GWTC-3: Compact Binary Coalescences Observed by LIGO and Virgo during the Second Part of the Third Observing Run}",
      journal = {Physical Review X},
         year = 2023,
        month = oct,
       volume = {13},
       number = {4},
          eid = {041039},
        pages = {041039},
          doi = {10.1103/PhysRevX.13.041039},
archivePrefix = {arXiv},
       eprint = {2111.03606},
 primaryClass = {gr-qc},
       adsurl = {https://ui.adsabs.harvard.edu/abs/2023PhRvX..13d1039A}
}

@article{knee22, 
       author = {{Knee}, Alan M. and {Romero-Shaw}, Isobel M. and {Lasky}, Paul D. and {McIver}, Jess and {Thrane}, Eric},
        title = "{A Rosetta Stone for Eccentric Gravitational Waveform Models}",
      journal = {The Astrophysical Journal},
         year = 2022,
        month = sep,
       volume = {936},
       number = {2},
          eid = {172},
        pages = {172},
          doi = {10.3847/1538-4357/ac8b02},
archivePrefix = {arXiv},
       eprint = {2207.14346},
 primaryClass = {gr-qc},
       adsurl = {https://ui.adsabs.harvard.edu/abs/2022ApJ...936..172K}
}

@article{romero_shaw_22,
       author = {{Romero-Shaw}, Isobel and {Lasky}, Paul D. and {Thrane}, Eric},
        title = "{Four Eccentric Mergers Increase the Evidence that LIGO-Virgo-KAGRA's Binary Black Holes Form Dynamically}",
      journal = {\apj},
         year = 2022,
        month = dec,
       volume = {940},
       number = {2},
          eid = {171},
        pages = {171},
          doi = {10.3847/1538-4357/ac9798},
archivePrefix = {arXiv},
       eprint = {2206.14695},
 primaryClass = {astro-ph.HE},
       adsurl = {https://ui.adsabs.harvard.edu/abs/2022ApJ...940..171R}
}

@ARTICLE{clarke_2022,
       author = {{Clarke}, Teagan A. and {Romero-Shaw}, Isobel M. and {Lasky}, Paul D. and {Thrane}, Eric},
        title = "{Gravitational-wave inference for eccentric binaries: the argument of periapsis}",
      journal = {\mnras},
         year = 2022,
        month = dec,
       volume = {517},
       number = {3},
        pages = {3778-3784},
          doi = {10.1093/mnras/stac2965},
archivePrefix = {arXiv},
       eprint = {2206.14006},
 primaryClass = {gr-qc},
       adsurl = {https://ui.adsabs.harvard.edu/abs/2022MNRAS.517.3778C}
}

@ARTICLE{Morras2025,
       author = {{Morras}, Gonzalo and {Pratten}, Geraint and {Schmidt}, Patricia},
        title = "{Orbital Eccentricity in a Neutron Star─Black Hole Merger}",
      journal = {The Astrophysical Journal},
         year = 2026,
        month = mar,
       volume = {1000},
       number = {1},
          eid = {L2},
        pages = {L2},
          doi = {10.3847/2041-8213/ae474c},
archivePrefix = {arXiv},
       eprint = {2503.15393},
 primaryClass = {astro-ph.HE},
       adsurl = {https://ui.adsabs.harvard.edu/abs/2026ApJ..1000L...2M}
}

@ARTICLE{Gamboa2024,
       author = {{Gamboa}, Aldo and {Buonanno}, Alessandra and {Enficiaud}, Raffi and {Khalil}, Mohammed and {Ramos-Buades}, Antoni and {Pompili}, Lorenzo and {Estell{\'e}s}, H{\'e}ctor and {Boyle}, Michael and {Kidder}, Lawrence E. and {Pfeiffer}, Harald P. and et al.},
        title = "{Accurate waveforms for eccentric, aligned-spin binary black holes: The multipolar effective-one-body model SEOBNRv5EHM}",
      journal = {Physical Review D},
         year = 2025,
        month = aug,
       volume = {112},
       number = {4},
          eid = {044038},
        pages = {044038},
          doi = {10.1103/jxrc-z298},
archivePrefix = {arXiv},
       eprint = {2412.12823},
 primaryClass = {gr-qc},
       adsurl = {https://ui.adsabs.harvard.edu/abs/2025PhRvD.112d4038G}
}

@ARTICLE{Ramos-Buades2023,
       author = {{Ramos-Buades}, Antoni and {Buonanno}, Alessandra and {Estell{\'e}s}, H{\'e}ctor and {Khalil}, Mohammed and {Mihaylov}, Deyan P. and {Ossokine}, Serguei and {Pompili}, Lorenzo and {Shiferaw}, Mahlet},
        title = "{Next generation of accurate and efficient multipolar precessing-spin effective-one-body waveforms for binary black holes}",
      journal = {\prd},
         year = 2023,
        month = dec,
       volume = {108},
       number = {12},
          eid = {124037},
        pages = {124037},
          doi = {10.1103/PhysRevD.108.124037},
archivePrefix = {arXiv},
       eprint = {2303.18046},
 primaryClass = {gr-qc},
       adsurl = {https://ui.adsabs.harvard.edu/abs/2023PhRvD.108l4037R}
}

@ARTICLE{Morras2025_waveform,
       author = {{Morras}, Gonzalo and {Pratten}, Geraint and {Schmidt}, Patricia},
        title = "{Improved post-Newtonian waveform model for inspiralling precessing-eccentric compact binaries}",
      journal = {Physical Review D},
         year = 2025,
        month = apr,
       volume = {111},
       number = {8},
          eid = {084052},
        pages = {084052},
          doi = {10.1103/PhysRevD.111.084052},
archivePrefix = {arXiv},
       eprint = {2502.03929},
 primaryClass = {gr-qc},
       adsurl = {https://ui.adsabs.harvard.edu/abs/2025PhRvD.111h4052M}
}

@ARTICLE{Xu2026,
       author = {{Xu}, Yumeng and {Valencia}, Jorge and {Estrella}, H{\'e}ctor Estell{\'e}s and {Ramos-Buades}, Antoni and {Husa}, Sascha and {Rossell{\'o}-Sastre}, Maria and {Querol}, Joan Llobera and {Ramis Vidal}, Felip A. and {Llompart}, Maria de Lluc Planas and {Colleoni}, Marta and et al.},
        title = "{Parameter estimation for the GWTC-4.0 catalog with phenomenological waveform models that include orbital eccentricity and an updated description of spin precession}",
      journal = {Physical Review D},
         year = 2026,
        month = apr,
       volume = {113},
       number = {8},
          eid = {083001},
        pages = {083001},
          doi = {10.1103/w6vy-v8vz},
archivePrefix = {arXiv},
       eprint = {2512.19513},
 primaryClass = {gr-qc},
       adsurl = {https://ui.adsabs.harvard.edu/abs/2026PhRvD.113h3001X}
}

@ARTICLE{Gupte2026,
       author = {{Gupte}, Nihar and {Miller}, M. Coleman and {Udall}, Rhiannon and {Bini}, Sophie and {Buonanno}, Alessandra and {Gair}, Jonathan and {Gamboa}, Aldo and {Pompili}, Lorenzo and {Ramos-Buades}, Antoni and {Dax}, Maximilian and et al.},
        title = "{Eccentricity constraints disfavor single-single capture in nuclear star clusters as the origin of all LIGO-Virgo-KAGRA binary black holes}",
      journal = {arXiv e-prints},
         year = 2026,
        month = mar,
          eid = {arXiv:2603.29019},
        pages = {arXiv:2603.29019},
          doi = {10.48550/arXiv.2603.29019},
archivePrefix = {arXiv},
       eprint = {2603.29019},
 primaryClass = {astro-ph.HE},
       adsurl = {https://ui.adsabs.harvard.edu/abs/2026arXiv260329019G}
}

@ARTICLE{Tibrewal2026,
       author = {{Tibrewal}, Snehal and {Zimmerman}, Aaron and {Lange}, Jacob and {Shoemaker}, Deirdre},
        title = "{Misinterpreting Spin Precession as Orbital Eccentricity in Gravitational-Wave Signals}",
      journal = {arXiv e-prints},
         year = 2026,
        month = jan,
          eid = {arXiv:2601.02260},
        pages = {arXiv:2601.02260},
          doi = {10.48550/arXiv.2601.02260},
archivePrefix = {arXiv},
       eprint = {2601.02260},
 primaryClass = {gr-qc},
       adsurl = {https://ui.adsabs.harvard.edu/abs/2026arXiv260102260T}
}

@ARTICLE{deLlucPlanas2025,
       author = {{Planas}, Maria de Lluc and {Ramos-Buades}, Antoni and {Garc{\'\i}a-Quir{\'o}s}, Cecilio and {Estell{\'e}s}, H{\'e}ctor and {Husa}, Sascha and {Haney}, Maria},
        title = "{Time-domain phenomenological multipolar waveforms for aligned-spin binary black holes in elliptical orbits}",
      journal = {Physical Review D},
         year = 2026,
        month = jan,
       volume = {113},
       number = {2},
          eid = {024006},
        pages = {024006},
          doi = {10.1103/wz3v-b151},
archivePrefix = {arXiv},
       eprint = {2503.13062},
 primaryClass = {gr-qc},
       adsurl = {https://ui.adsabs.harvard.edu/abs/2026PhRvD.113b4006P}
}

@ARTICLE{Iglesias2024,
       author = {{Iglesias}, H.~L. and {Lange}, J. and {Bartos}, I. and {Bhaumik}, S. and {Gamba}, R. and {Gayathri}, V. and {Jan}, A. and {Nowicki}, R. and {O'Shaughnessy}, R. and {Shoemaker}, D.~M. and et al.},
        title = "{Eccentricity Estimation for Five Binary Black Hole Mergers with Higher-order Gravitational-wave Modes}",
      journal = {\apj},
         year = 2024,
        month = sep,
       volume = {972},
       number = {1},
          eid = {65},
        pages = {65},
          doi = {10.3847/1538-4357/ad5ff6},
archivePrefix = {arXiv},
       eprint = {2208.01766},
 primaryClass = {gr-qc},
       adsurl = {https://ui.adsabs.harvard.edu/abs/2024ApJ...972...65I}
}

@article{Shaikh2025,
author = "Shaikh, Md Arif and Varma, Vijay and Ramos-Buades, Antoni and Pfeiffer, Harald P. and Boyle, Michael and Kidder, Lawrence E. and Scheel, Mark A.",
title = "{Defining eccentricity for spin-precessing binaries}",
eprint = "2507.08345",
archivePrefix = "arXiv",
primaryClass = "gr-qc",
doi = "10.1088/1361-6382/ae085d",
journal = "Class. Quant. Grav.",
volume = "42",
number = "19",
pages = "195012",
year = "2025",
note = "{\href{https://pypi.org/project/gw_eccentricity}{pypi.org/project/gw\_eccentricity}}",
}

@ARTICLE{Malagon2026,
       author = {{Malagon}, Natalie and {O'Shaughnessy}, Richard},
        title = "{Assessing the imprint of eccentricity in GW signatures using two independent waveform models}",
      journal = {arXiv e-prints},
         year = 2026,
        month = may,
          eid = {arXiv:2605.12818},
        pages = {arXiv:2605.12818},
archivePrefix = {arXiv},
       eprint = {2605.12818},
 primaryClass = {astro-ph.HE},
       adsurl = {https://ui.adsabs.harvard.edu/abs/2026arXiv260512818M}
}

\end{document}